\documentclass[preprint]{revtex4-1}
\usepackage{amsmath,graphicx,amssymb,fancyhdr,amsthm,textcomp}  
\usepackage{enumerate}
\usepackage[usenames]{color}

\theoremstyle{definition}
	
\theoremstyle{remark}

\usepackage{color}

\def\beq{\begin{eqnarray}}
\def\eeq{\end{eqnarray}}
\def\bsp{\begin{split}}
\def\esp{\end{split}}

\newcommand{\la}{{\lambda}}

\newcommand{\Csf}{{\sf C}}

\def\beq{\begin{eqnarray}}
\def\eeq{\end{eqnarray}}

\begin{document}

\begin{abstract} 

In this paper we introduce an algorithm to determine the equivalence of five dimensional spacetimes, which generalizes the Karlhede algorithm for four dimensional general relativity. As an alternative to the Petrov type classification, we employ the alignment classification to algebraically classify the Weyl tensor. To illustrate the algorithm we discuss three examples: the singly rotating Myers-Perry solution, the Kerr (anti) de Sitter solution, and the rotating black ring solution. We briefly discuss some applications of the Cartan algorithm in five dimensions.


\end{abstract}

\title{\Large\textbf{The Cartan Algorithm in Five
Dimensions}}
\author{\large\textbf{D. D. McNutt}}
\affiliation{Faculty of Science and Technology,\\ 
                         University of Stavanger, 
                         N-4036 Stavanger, Norway         }
\email{david.d.mcnutt@uis.no}
\author{\large\textbf{A. A. Coley}}
\affiliation{ Department of Mathematics and Statistics,\\ 
                         Dalhousie University, 
                         Halifax, Nova Scotia,\\ 
                         Canada B3H 3J5 }
\email{aac@mathstat.dal.ca, Adam.AL.Forget@dal.ca}
\author{\large\textbf{A. Forget}}
\affiliation{ Department of Mathematics and Statistics,\\ 
                         Dalhousie University, 
                         Halifax, Nova Scotia,\\ 
                         Canada B3H 3J5 }
\date{\today}   
\maketitle   
\pagestyle{fancy}   
\fancyhead{} 
\fancyhead[EC]{A. A.~Coley and D. D. McNutt}   
\fancyhead[EL,OR]{\thepage}   
\fancyhead[OC]{}   
\fancyfoot{} 


\section{Introduction}

There are significant differences between higher dimensional spacetimes in general relativity (GR), and their analogues in four dimensions (4D). Gravity in higher dimensions exhibits a much richer mathematical structure than in 4D. One important example arises with black hole solutions: in 4D GR, the Kerr black hole is unique, while in higher dimensions there exist a number of different asymptotically flat, higher-dimensional vacuum black hole solutions \cite{HIGHER-D-REVIEW}. This prompts the question of the classification of higher dimensional solutions and deciding when any two spacetimes are equivalent. 

Two Lorentzian manifolds, $(M, g)$ and $(\bar{M},\bar{g})$ are equivalent if there exists a  locally-defined diffeomorphism $\Phi:M\rightarrow \bar{M}$ between them such that 
$$\Phi^*(\bar{g})=g.$$ 
\noindent The scalar polynomial curvature invariants (SPIs) may be used to show inequivalence of spacetimes. However, SPIs are not sufficient to prove equivalence of two spacetimes.  \'Elie Cartan developed an approach for determining the equivalence of sets of differential forms defined on differentiable manifolds under appropriate transformation groups \cite{kramer, olver}. 

In 4D, the Karlhede algorithm is an adaptation of the Cartan algorithm to the special case of Lorentzian metrics. \cite{kramer}. The classification of Lorentzian metrics of Petrov type {\bf D}  has been resolved in 4D \cite{ref1}, and so any 4D  type {\bf D} black hole solution can be classified. The differences between the isotropy groups of the 4D and higher dimensional Lorentzian metrics have hindered the development of a higher dimensional analogue of the Karlhede algorithm, but such a classification is still possible. In this paper we will introduce an adaptation of the Cartan algorithm to five dimensional (5D) spacetimes and apply this to three important black hole solutions.

\subsection{The Cartan Algorithm}

To compare two different metrics at each point on their respective manifolds, we examine their coordinate neighbourhoods and consider coordinate or basis transformations. The issue of equivalence is decided on the frame bundle of the manifold: if the metrics are equivalent, the frame bundles derived from them will be (locally) identical \cite{kramer}. This is accomplished by fixing the curvature tensor and its covariant derivatives using frame transformations and recording those frame transformations which do not change the form of these tensors \cite{olver}

The algorithmic procedure for determining equivalence is as follows:

\begin{enumerate}
\item Set $q$, the order of differentiation, to 0.
\item Compute up to the the $q^{th}$ covariant derivatives of the Riemann tensor.
\item Fix the Riemann tensor and its covariant derivatives in a canonical form.
\item Fix the frame as much as possible using this canonical form, and record the remaining frame freedom (the group of allowed transformations is the {\it linear isotropy group $H_q$}). 
\item Find the number $t_q$ of independent functions in the components of the Riemann tensor and its covariant derivatives, in the canonical form.
\item If the number of independent functions, and the isotropy group  are the same as in the previous step, let $p+1=q$, and the algorithm terminates; if they differ (or if $q=0$), increase $q$ by 1 and go to step 2. 
\end{enumerate}


\noindent The components of the Riemann tensor and its covariant derivatives relative to the frame fixed at the end of the Cartan algorithm are called {\it Cartan invariants}. 

The D-dimensional space-time is characterized by the sequences of isotropy groups and number of functionally independent invariants at each order, the canonical forms used, and the the Cartan invariants themselves. As there are $t_p$ essential spacetime coordinates, the remaining $D-t_p$ are ignorable, implying the dimension of the isotropy group of the space-time will be $s=\dim(H_p)$, and the isometry group has dimension $r=s+D-t_p$. To compare two space-times one can first compare these discrete invariants. If they match for each metric, we must compare the forms of the Cartan invariants relative to the same frame to determine equivalence.


\subsection{Algebraic Classification of Vacuum Spacetimes}

To fix the Riemann tensor and its covariant derivatives in a canonical form at each iteration of the algorithm, it will be useful to consider  the algebraic classification of tensors, and in particular the Riemann tensor. The algebraic classification of spacetimes has played a crucial role in the understanding of 4D solutions \cite{kramer}. We examine vacuum solutions where the Ricci tensor either vanishes or is proportional to the metric tensor, and so we will concern ourselves with the classification of the Weyl tensor. 

In 4D, algebraic classification can be accomplished in several different ways, using null vectors, 2-spinors, or bivectors (or even scalar invariants). Each of these can be applied to give a different description (or part) of the 4D algebraic classification scheme. In higher dimensions, algebraic classification may be generalized using each of these methods \cite{class, BIVECTOR}; however, each approach leads to distinct classification \cite{BIVECTOR,spinorclass}. 
The most well-studied approach providing an inclusive classification \cite{class,Alignment,AlignmentReview} examines the behaviour of the components of the Weyl tensor relative to a null frame under local Lorentz boosts. That is, the classification relies on a null frame: 
\beq\label{vecbasis}
  \{\ell,n,m_i\}, \qquad\qquad i=2,3,\ldots,D-1 
\eeq
where $\ell$ and $n$ are linearly independent null vectors, transforming as
\beq
  \ell \mapsto \la \ell, \qquad n\mapsto \la^{-1} n, \qquad m_i \mapsto m_i
\eeq
under a Lorentz boost where $\lambda$ is real-valued. Relative to the basis $\{ \theta^a\} = \{ \ell, n, m^i\}$, the components of an arbitrary tensor of rank $p$ transforms under a boost:
\beq T'_{a_1 a_2...a_p} = \lambda^{b_{a_1 a_2 ... a_p}} T_{a_1 a_2 ... a_p},~~ b_{a_1 a_2...a_p} = \sum_{i=1}^P(\delta_{a_i 0} - \delta_{a_i 1}) \nonumber \eeq
\noindent where $\delta_{ab}$ denotes the Kronecker delta symbol.  The quantity $b_{a_1 \cdots a_p}$ is called the {\it boost weight} (b.w) of the frame component $T_{a_1 a_2 ... a_p}$. We say that a null direction $\ell$ is {\it aligned} with a tensor $T$ if the components with the largest weight vanishes along that direction. The classification of a tensor using its b.w. will be referred to as the \emph{alignment classification}. 

The concept of alignment can be made more rigorous by defining the {\it boost order} of a tensor ${\bf T}$ as the maximum $b_{a_1 \cdots a_p}$ for all non-vanishing $T_{a_1 \cdots a_p}$; this quantity is dependent on the null direction only $\ell$, and hence we denote the boost order of a tensor ${\bf T}$ as $\mathcal{B}(\ell)$. We say that a null vector, $\ell$, is {\it aligned} with the Weyl tensor whenever $\mathcal{B}(\ell) \leq 1$ and call this vector, $\ell$, a {\it Weyl aligned null direction} (WAND). Spacetimes may be divided into six different primary types \cite{class,AlignmentReview,PraPraOrt07,higherghp}:  {\bf G}, {\bf I}, {\bf II}, {\bf III}, {\bf N}, {\bf O}  if there exist an aligned $\ell$ with $\mathcal{B}(\ell) = 1,0,-1,-2$. If in the case of type {\bf II}, there are multiple distinct WANDs, then this is of type {\bf D}. 

In 4D the alignment classification reproduces the Petrov classification; however, the algebraic types defined by the higher-dimensional alignment classification are quite broad in comparison to the 4D case \cite{BIVECTOR, CH2011,CHOL2012}. This classification can be refined using the higher-dimensional bivector classification which analyses the bivector map,
\beq\label{map:bivector}
 \Csf: X_{\mu\nu} \mapsto \tfrac{1}{2} C^{\phantom{\mu\nu}\rho\sigma}_{\mu\nu} X_{\rho\sigma}.
\eeq
By defining the bivector operator to be consistent with the b.w. decomposition, the components of fixed b.w. may be characterized in terms of basic constituents which transform under irreducible representations of the spins. This refinement to the alignment classification relies on the geometric relations between the highest b.w. constituents, and hence will be called the {\it spin type} \cite{BIVECTOR}. In 4D, considering the spin weight does not refine the alignment classification, which is equivalent to the Petrov classification.

To determine the alignment classification for a particular spacetime one must identify  WANDs. In four dimensions this could be achieved by applying null rotations and solving the resulting polynomial equations for the null rotation parameters; however, in higher dimensions this approach is infeasible. As an alternative one may use the generalization of the 4D Bel-Debever criteria to higher dimensions \cite{bdcond}: 
\beq & \ell^a \ell^c \ell_{[c} C_{a]bc[d}\ell_{f]} = 0 \leftarrow \ell \text{ is a WAND, at most primary type {\bf I}}. \label{TypeIcond} &  \\
& \ell^b \ell^c  C_{abc[d}\ell_{e]} = 0 \leftarrow \ell \text{ is a WAND, at most primary type {\bf II}}.  & \label{TypeIIcond} \\
& \ell^c  C_{abc[d}\ell_{e]} = 0 \leftarrow \ell \text{ is a WAND, at most primary type {\bf III}}.  &  \\
& \ell^c  C_{abcd} = 0 \leftarrow \ell \text{ is a WAND, at most primary type {\bf N}},  &  \eeq

\noindent to determine $\ell$. If multiple distinct WANDs exist for which the Weyl tensor is of type {\bf II} then this is of type {\bf D}. The singly-rotating Myers Perry and Kerr-(Anti) de Sitter black holes are of type {\bf D}. The rotating black ring admits regions where it is of type ${\bf I}_i$ (i.e., of type ${\bf I}$ for both $\ell$ and $n$) or more special, and regions where it is of type ${\bf I}/ {\bf G}$.


Due to the differing algebraic type of the Weyl tensor, the rotating black ring (RBR) and Kerr-(Anti) de Sitter (Kerr-(A)dS) black hole are not equivalent. However, they both specialize to the singly rotating Myers Perry black hole by choosing the parameters appropriately. This will be reflected in the structure of the Cartan invariants for each spacetime, and from this conclusion, one could imagine a general pair of WANDs where 
$$ \hat{L}_{\pm} =  L_{ MP, \pm} + \xi L'_{Kerr-(A)dS, \pm} + \chi L'_{RBR, \pm} \nonumber  $$ 
\noindent where the vanishing of $\xi$ or $\chi$ produce the WANDs for the Kerr-(A)dS or RBR metrics respectively.    
\newpage
\section{Five dimensional Spacetimes}

In higher dimensions, the Petrov classification can not be implemented as in 4D. The Weyl tensor no longer has a dual of the same rank, and thus we cannot build an operator that acts on the space of self-dual bivectors. However, we are able to use the alignment classification to algebraically classify the Weyl tensor, and when possible employ WANDs to put the Weyl tensor into a simpler form. 

For certain spacetimes the identification of WANDs may be difficult to determine, and when the determination of the WANDs is computationally infeasible we can employ an alternative choice of coframe to continue the Cartan algorithm. The bivector classification suggests alternative canonical forms relative to the isotropy group of the Weyl tensor \cite{BIVECTOR}.

The alignment classification for 5D spacetimes can be made finer by considering the spin types arising from the spin group (which is isomorphic to $O(3)$) acts on the null frame according to \cite{CHOL2012}: \beq \ell' = \ell ,~~n' = n,~~m^{i'} = m^j X_j^{i}. \nonumber \eeq \noindent This approach can be used to identify the isotropy group at the zeroth and first iteration. 


\subsection{5D Lorentz transformation}

When applying the Cartan algorithm  we must choose a canonical form for the curvature tensor by fixing the
frame using the Lorentz transformations. We will consider the action of the Lorentz transformation case by case. To define these we introduce a coframe consisting of two null vectors and three spacelike vectors:
\beq
l_a l^a=n_a n^a=0, \quad l_a n^a=1, \quad m_{ia}m^{ja}=\delta_i^{~j}, 
\eeq
such that the metric reads as
\beq
g_{ab}=-2l_{(a}n_{b)}+\delta_{ij}m^i_{~a}m^j_{~b}\,,
\eeq
with round parentheses denoting symmetrization. 

In terms of this frame basis the Lorentz transformation are defined by (in 5D)
\cite{class,Alignment}:

\begin{itemize}
  \item Null rotations: $\hat l= l+z_i m^i+\frac14 z_i z^i n, \quad \hat n= n, \quad \hat m_i= m_i+z_i n $
  \item Boost: $\hat l= \lambda l , \quad \hat n=\lambda^{-1} n , \quad \hat m_i = m_i\,.$
  \item Spins: $\hat l=l , \quad \hat n=n , \quad \hat m_i =X_i^j m_j. $  where $X_i^j$ denotes the usual rotation matrices (about $m_1$, $m_2$, $m_3$, respectively)


\end{itemize}
\noindent The quantities $z_i=z_i(x^a)$, $X_i^j=X_i^j(x^a)$ and $\lambda=\lambda(x^a)$ are real-valued functions of the coordinates.

\subsection{Refinement of the Weyl Tensor Classification in 5D}  \label{sec:refine}

The indicial symmetries of the Weyl tensor imply that all components of b.w. $\pm 4$ or $\pm 3$ are zero, and that the remaining components of fixed b.w. satisfy algebraic relations:
\beq & b.w.~ 2:~C_{0~0i}^{~i} = 0;,~~~~ b.w.~-2:~C_{1~1i}^{~i} = 0& \nonumber \\
&b.w.~1:~C_{010i} = C_{0~ij}^{~j};,~~~~ b.w.~-1:~C_{101i} = C_{1~ij}^{~j} & \nonumber \\
&b.w.~0:~2C_{0(ij)1} = C_{i~jk}^{~k},~~~~2C_{0[ij]1} = - C_{01ij},~~~~2C_{0101} = - C^{ij}_{~~ij} = 2C_{0~1i}^{~i}. & \nonumber \eeq

\noindent The independent Weyl tensor components of a fixed b.w. $q$ define objects which transform under irreducible representations of the spin group, these are called the {\it Weyl constituents}. 

For a general spacetime of dimension $D=n+2$,  the b.w. 2 and -2 components are already in the appropriate form
\beq \hat{H} = C_{0i0j},~~ \check{H} = C_{1i1j}. \nonumber \eeq
\noindent The b.w. 0 b.w. -1 components $C_{ijkl}$ and $C_{1ijk}$ may be decomposed as \cite{BIVECTOR,CHOL2012}:
\beq C^{ij}_{~~kl} &=& \bar{H}^{[ij]}_{~~~[kl]} = \bar{C}^{ij}_{~~kl} + \frac{4}{n-2} \delta^{[i}_{~[k} \bar{S}^{j]}_{~l]} + \frac{2}{n(n-1)} \bar{R} \delta^{[i}_{~[k} \delta^{j]}_{~l]} \nonumber \\
C_{1ijk} &=& \check{L}_{i[jk]} = 2 \delta_{i[j} \check{v}_{k]} + \check{T}_{ijk},~~~ \check{T}^i_{~ik} = \check{T}_{i(jk)} =0 \nonumber \eeq
\noindent where $\bar{H}_{ijkl}$ is a n-dimensional Riemann-like tensor, and $\bar{C}_{ijkl}, \bar{R}= \bar{H}^{ij}_{~~ij}$, and $\bar{S}_{ij} = \bar{H}^k_{~ikj}-\frac{1}{n}\bar{R} \delta_{ij}$ are the associated Weyl tensor, Ricci scalar and tracefree Ricci tensor respectively. 

In 5D, the transverse space is 3-dimensional (3D) and so $\bar{C}_{ijkl} = 0$, while the b.w. -1 constituent $\check{T}$ is equivalent to a traceless symmetric matrix $\check{n}$ using the alternating Levi-Civita symbol in 3D:
\beq \check{n}_{ij} = \frac{1}{2} \epsilon^{kl}_{~~(i} \check{T}_{j)kl},~~ \check{n}_{ij} = \check{n}_{(ij)},~~ \check{n}^i_{~i} =0 \nonumber \eeq
\noindent In a similar manner, the component $C_{0ijk}$ give rise to the b.w. 1 constituents $\hat{v}$ and $\hat{n}$. In 5D the remaining b.w. 0 component, $C_{01ij} = A_{ij}$ is an anti-symmetric matrix in the 3D transverse space, and so we may work with its dual vector $\bar{w}$:
\beq \bar{w}_i =  \frac{1}{2} \epsilon_{ijk} A^{jk} \nonumber \eeq
These results may be summarized for the 5D Weyl tensor in the following table. For a given algebraic type, we can apply the spatial rotations to simplify the form of the constituent quantities, and put the Weyl tensor into a canonical form. In the case that the constituents of the Weyl tensor are vectors and matrices, this provides important geometric information that can be used for the Cartan algorithm.

\begin{table}[b] \label{table:1}
\beq \begin{array}{|c|c|c|} \hline
 \text{ b.w. } & \text{ Constituents } & \text{ Weyl tensor Components} \\ \hline
+2 & \hat{H}_{ij} & C_{0i0j} = \hat{H}_{ij} \\ \hline
+1 & \hat{n}_{ij},~\hat{v}_i & C_{0ijk} = 2\delta_{i[j} \hat{v}_{k]}+ \hat{n}_i^{~l}\epsilon_{ljk} \\ 
 & & C_{010i} = -2\hat{v}_i \\ \hline
 0 & \bar{S}_{ij},~~ \bar{w}_i,~~\bar{R} & C^{ij}_{~~kl} = 4 \delta^{[i}_{~~[k} \bar{S}^{j]}_{~~l]} + \frac13 \bar{R} \delta^{[i}_{~~[k} \delta^{j]}_{~~l]} \\ 
 & & C_{1i0j} = M_{ij} = - \frac12 \bar{S}_{ij} - \frac16 \bar{R} \delta_{ij} - \frac12 \epsilon_{ijk} \bar{w}^k \\
 & & C_{01ij} = A_{ij} = \epsilon_{ijk} \bar{w}^k \\
 & & C_{0101} = -\frac12 \bar{R} \\ \hline
-1 & \check{n}_{ij},~\check{v}_i & C_{1ijk} = 2\delta_{i[j} \check{ v}_{k]}+ \check{n}_i^{~l}\epsilon_{ljk} \\
 & & C_{101i} = -2\check{ v}_i \\ \hline
-2 & \check{H}_{ij} & C_{1i1j} = \check{H}_{ij} \\ \hline
\end{array}
\nonumber \eeq
\caption{Constituent parts of the 5D Weyl tensor \cite{CHOL2012}. Here $\epsilon_{ijk}$ is the alternating Levi-Civita symbol for the 3D transverse space.}
\end{table}

\section{Examples}

In this section we review three black hole solutions in 5D. We discuss the singly-rotating Myers-Perry black hole, the Kerr- (Anti-) de Sitter black hole, and the rotating black ring. The first two solutions are of type {\bf D}, while the last is of type ${\bf I}_i$.

\subsection{Singly-Rotating Myers-Perry}
The singly-rotating Myers-Perry metric is a 5D analogue of the Kerr metric. Choosing coordinates $(t,x,y,\phi, \psi)$ with $-1 < x < 1$, and $\Delta \phi = \Delta \Psi = \frac{2\sqrt{2} \pi}{1+\nu}$ (i.e., they are periodic with equal period), and $y \in (-\infty, 1]$ or $y \in [1/\nu, \infty)$ \cite{EE:2003}, the metric is then: 
\beq
\mathrm{d}s^2 &= -\sqrt{\frac{1-x}{1-y}}(\mathrm{d}t+R\sqrt{\nu}(1+y)\mathrm{d}\psi)^2+\frac{R^2}{(x-y)^2}[(x-1)((1-y^2)(1-\nu y)\mathrm{d}\psi^2 \nonumber \\
	&+ \frac{\mathrm{d}y^2}{(1+y)(1-\nu y)})+(1-x)^2(\frac{\mathrm{d}x^2}{(1-x^2)(1-\nu x)} + (1+x)(1-\nu x)\mathrm{d}\phi^2)].
\eeq
\noindent The parameter $R$ acts as a length scale, while $\nu$ is a dimensionless rotation parameter.

We define a non-normalized null frame $\{L_+, L_-, \partial_\phi, \partial_y, \partial_\psi \}$ using WANDs defined in \cite{brwands} and which both satisfy \eqref{TypeIIcond} relative to the coordinate basis:
\beq
L_\pm &=& \frac{1}{(x^2 -1)(\nu y-1)} \left(\frac{\nu yx-y+\nu x+1-2\nu y}{x-y}R\partial_t -\sqrt{\nu}\partial_\psi \right) \nonumber \\ && \pm \sqrt{\frac{\nu x-1}{(x-y)(y-1)}} \left(\partial_x + \frac{y^2-1}{x^2-1}\partial_y \right).
\eeq

\noindent The pair of null vectors are not unique, as any boost will generate a new pair of WANDs. Employing the Gram-Schmidt procedure we then build a normalized non-coordinate null frame $\{l, n, m^2, m^3, m^4\}$ with $l\sim L_+$ and $n\sim L_-$. Relative to this frame, the only non-zero components of the Weyl tensor are those with b.w. zero.

At zeroth order the Weyl tensor is of type D, using the decomposition of the Weyl tensor in table I, the algebraically independent components can be expressed as a scalar and two tensors: 

\beq R = 2 C_{0101} = \frac{2(x-y)^2(4\nu x+\nu -3)}{4(y-1)^2 R^2} \eeq



\beq
A_{ij} = C_{01ij} =
\begin{pmatrix}
0 & \frac{\sqrt{(1-\nu x)(\nu )(x+1)}(x-y)^2}{(y-1)^2 R^2} & 0 \\
-\frac{\sqrt{(1-\nu x)(\nu )(x+1)}(x-y)^2}{(y-1)^2 R^2} & 0 & 0 \\
0 & 0 & 0
\end{pmatrix}
\eeq


\small
\beq
M_{ij} = C_{1i0j}  =
\begin{pmatrix}
-\frac{1}{4} \frac{(x-y)^2 (\nu -1)}{(y-1)^2 R^2} & -\frac{1}{2} \frac{\sqrt{(1-\nu x)(\nu )(x+1)}(x-y)^2}{(y-1)^2 R^2} & 0 \\
\frac{1}{2} \frac{\sqrt{(1-\nu x)(\nu )(x+1)}(x-y)^2}{(y-1)^2 R^2} & \frac{1}{4} \frac{(x-y)^2 (2\nu x+\nu -1)}{(y-1)^2 R^2} & 0 \\
0 & 0 & \frac{1}{4} \frac{(x-y)^2 (2\nu x+\nu -1)}{(y-1)^2 R^2}
\end{pmatrix}
\eeq
\normalsize
\noindent Noting that only two coordinates appear in the above functions, it can be shown that the components of the Weyl tensor are functionally dependent on any two components (say, for example, $C_{1010}$ and $C_{0123}$) and hence $t_0 = 2$.  


As all non-zero b.w. terms vanish, the Weyl tensor is invariant under a boost. Since $A_{ij}=\epsilon_{ijk} w^k$, and the vector ${\bf \bar{w}}$, with components  $\bar{w}^k$, is proportional to $m_4$, rotations about $m^4$ will not affect $A_{ij}$. Similarly one can show that $M_{ij}$ is invariant under a rotation about $m^4$. To verify if this rotation is an isotropy of the Weyl tensor we check that $S_{ij}$ is unchanged under a rotation about $m^4$. Since $\frac12 S_{ij}=-M_{ij}-\frac{1}{6}R\delta_{ij}-\frac{1}{2}A_{ij}$ it follows that $\hat{S}_{ij}$ is unfaffected by the rotation. Therefore, rotations about $m^4$ belong to the zeroth order isotropy group and $\dim(H_0)=2$.

For the first iteration of the Karlhede algorithm, we obtain a large list of non-zero components of the first covariant derivative of the Weyl tensor. There are too many to show here, but we display two non-zero invariants: 
\beq
& C_{1010;0} = -\frac{3\sqrt{2}}{4} \frac{(2\nu x+\nu-1)(x-y)^{5/2} \sqrt{(1-\nu x)(\nu y -1)(x-1)}}{\sqrt{\nu +1}(y-1)^3 R^3} & \\
& C_{1010;2} = -\frac38 \frac{(3\nu  x + 2 \nu -1)(\nu x-1)(x-y)^5(2\nu x + v -1)(-1+x(\nu y -1))}{(y-1)^6(\nu +1) R^6}; &
\eeq

\noindent these are notable as they show that we may fix the  boost parameter to set $C_{1010;0} = 1$, and set the parameter for a rotation about $m^4$ to zero by fixing $C_{1010;2} \neq 0$ and $C_{1010;3} = 0$. This implies the isotropy group at the first iteration is zero, $dim(H_1) = 0$. It can be shown that the components of the covariant derivative of the Weyl tensor are functionally dependent of any two zeroth order invariants found earlier, implying that $t_1 = 2$.  

The Cartan algorithm must continue to the second iteration to conclude the algorithm; no new functionally independent invariants appear nor can the isotropy group be reduced any further, thus $t_2 = 2$ and $\dim(H_2)=0$, and the algorithm terminates. The second order Cartan invariants are important for the classification of the spacetime, but we will not display them here. 

\subsection{Kerr-ADS  Metric}

In Kerr-Schild form, the 5D Kerr-de Sitter metric will be $ds^2 = d\bar{s}^2 +\frac{2M}{\rho^2}(k_\mu dx^\mu)^2$ where the de Sitter metric is given by:
\beq d\bar{s}^2 &=& - \frac{(1-\lambda r^2) \Delta dt^2}{(1+\lambda a^2)(1+\lambda b^2)}+ \frac{r^2 \rho^2 dr^2}{(1-\lambda r^2)(r^2 +a^2)(r^2+b^2)} + \frac{\rho^2 d\theta^2}{\Delta}  \nonumber \\
&& + \frac{r^2+a^2}{1+\lambda a^2}\sin^2 \theta d\phi^2 + \frac{r^2+b^2}{1+\lambda b^2} \cos^2 \theta d\psi^2, \label{dSmetric} \eeq
\noindent with functions:
\beq \rho^2 = r^2 + a^2 \cos^2 \theta + b^2 \sin^2 \theta,~~\Delta = 1 +\lambda a^2 \cos^2 \theta + \lambda b^2 \sin^2 \theta, \nonumber \eeq

\noindent and the null vector is given by
\beq k_\mu dx^\mu = \frac{\Delta dt}{(1+\lambda a^2)(1+\lambda b^2)} + \frac{r^2 \rho^2 dr}{(1-\lambda r^2)(r^2 + a^2)(r^2+b^2)} - \frac{a \sin^2 \theta d\phi}{1+\lambda a^2} - \frac{b \cos^2 \theta d\psi }{1+\lambda b^2}. \nonumber \eeq
From  \cite{CP2005} we consider the null coframe 
\beq & \ell = k,~~ n = A dt + B dr + J d\phi + K d\psi,& \nonumber \\ & m_2 = \frac{\rho}{\sqrt{\Delta}} d\theta,~~m_3 = H dt + F d\phi,~~ m_4 = W dt + Z d\phi + X d\psi & \nonumber \eeq 
where: 
\beq & A = \frac{\Delta (2Mr^2 - R)}{2r^2 \rho^2(1+\lambda a^2)(1+ \lambda b^2)},~~B = \frac12 + \frac{Mr^2}{R},~~J = - \frac{a\sin^2 \theta (2Mr^2 - R)}{2r^2 \rho^2 (1+\lambda a^2)},~~ K = - \frac{b \cos^2 \theta (2Mr^2 - R)}{2r^2 \rho^2 (1+\lambda b^2)} & \nonumber \\
& H = - \frac{\sqrt{\Delta} (1-\lambda r^2) a \sin \theta}{(1+\lambda a^2) \sqrt{S}},~~ F = \frac{\sqrt{\Delta} (r^2+a^2) \sin \theta}{(1+\lambda a^2) \sqrt{S}},~~ W = -\frac{\Delta (r^2 +a^2) (1-\lambda r^2) b \cos \theta}{r\rho (1+\lambda a^2)(1+\lambda b^2) \sqrt{S}}& \nonumber \\
& Z = \frac{(r^2+a^2)(1-\lambda r^2) ab \sin^2 \theta \cos \theta}{r \rho (1 + \lambda a^2) \sqrt{S}},~~X = \frac{(r^2 + b^2) \cos \theta \sqrt{S}}{r\rho (1+\lambda b^2)}, \nonumber \eeq
\noindent and we have defined $R = (r^2+a^2)(r^2+b^2)(1-\lambda r^2)$ and $S = \rho^2 - (1-\lambda r^2) b^2 \sin^2 \theta$. 

Relative to this coframe the Weyl tensor is of type {\bf D}. To write the constituent components in a simpler form, we apply a spatial rotation about $m^2$ with the parameter defined as:

\beq \tan(z) = \frac{\sin \theta  b \sqrt{1 + \lambda b \sin^2 \theta + \lambda a^2 \cos^2 \theta } r (r^2 + b^2 \sin^2 \theta + a^2 \cos^2 \theta)^{\frac52}}{(-r^2 -b^2 \sin^2 \theta - a^2 \cos^2 \theta)^3 a \cos \theta}. \nonumber \eeq

The scalar $R$ and matrices $A_{ij}$ and $M_{ij}$ are then 

\beq R = \frac{2M (- a^2 \cos^2 \theta - b^2 \sin^2 \theta + 3 r^2  )}{(-r^2 - b^2 sin^2 \theta - a^2 \cos^2 \theta )^3} \nonumber \eeq

\beq 
A_{ij} =
\begin{pmatrix}
0 & -\frac{8 \sqrt{a^2 \cos^2 \theta + b^2 \sin^2 \theta} M r}{(r^2+a^2 \cos^2 \theta + b^2 \sin^2 \theta)^3} & 0 \\
\frac{8 \sqrt{a^2 \cos^2 \theta + b^2 \sin^2 \theta} M r}{(r^2+a^2 \cos^2 \theta + b^2 \sin^2 \theta)^3} & 0 & 0 \\
0 & 0 & 0
\end{pmatrix}
\eeq
\beq
M_{ij} =
\begin{pmatrix}
-\frac{2M}{\rho^4  } & -\frac{8 \sqrt{a^2 \cos^2 \theta + b^2 \sin^2 \theta} M r}{(r^2+a^2 \cos^2 \theta + b^2 \sin^2 \theta)^3} & 0 \\
\frac{8 \sqrt{a^2 \cos^2 \theta + b^2 \sin^2 \theta} M r}{(r^2+a^2 \cos^2 \theta + b^2 \sin^2 \theta)^3}& -\frac{2M}{\rho^4  } & 0 \\
0 & 0 & -\frac{2M}{\rho^4  }
\end{pmatrix} 
\eeq
%
%

\noindent Using the results of \cite{CHOL2012}, we conclude that at the zeroth iteration, boosts and rotations around $m_4$ do not affect the Weyl constituents $R, A_{ij}, M_{ij}$ and $S_{ij}$; thus, the Weyl tensor is unchanged under boosts and rotations about $m_4$, and $dim(H_0) = 2$. By inspection we find that there are two functionally independent invariants, implying $t_0 = 2 $.

For the first iteration of the Karlhede algorithm, we obtain many non-zero components of the first covariant derivative of the Weyl tensor which will be omitted except for two non-zero invariants: 
\beq
& C_{1010;0} = \frac{24 M r (r^2 - a^2 \cos^2 \theta - b^2 \sin^2 \theta) }{(r^2 + a^2 \cos^2 \theta + b^2 \sin^2 \theta)^4} & \\
& C_{1010;2} = - \frac{8 M \sin \theta \cos \theta \sqrt{1+\lambda a^2 \cos^2 \theta + \lambda b^2 \sin^2 \theta} (a^2 - b^2)(5 r^2 - a^2 \cos^2 \theta - b^2 \sin^2 \theta)}{\sqrt{r^2 a^2 \cos^2 \theta + b^2 \sin^2 \theta}(r^2 - a^2 \cos^2 \theta - b^2 \sin^2 \theta)^4}. &
\eeq
\noindent  As in the case of the Myers-Perry metric, the parameters for a boost and a rotation about $m^4$ may be fixed so that $C_{1010;0} = 1$ and $C_{1010;2} \neq 0$ with $C_{1010;3} = 0$. Thus, the isotropy group at the first iteration is zero,  $dim(H_1) = 0$. Similarly, $t_1 = 2$, since it can be shown that the components of the covariant derivative of the Weyl tensor are functionally dependent of any two zeroth order invariants found at zeroth order.

The Cartan algorithm requires that we continue to second order to conclude the algorithm; no new functionally independent invariants appear nor can the isotropy group be reduced any further, thus $t_2 = 2$ and $\dim(H_2)=0$, and the algorithm terminates.

\subsection{Rotating Black Ring  Metric}

The rotating black ring (RBR) solution was found in \cite{RBR}; this solution has an interesting horizon topology $ S^1 \times S^2 $, which cannot occur in 4D. We will use the coordinates $(t,x,y,\phi, \psi)$ introduced in \cite{EE:2003} for the metric: 

\beq ds^2 &=& - \frac{F(x)}{F(y)} ( dt + R\sqrt{\lambda \nu} (1+y)d\psi)^2 \nonumber \\
&&+ \frac{R^2}{(x-y)^2} \left[ - F(x) \left( G(y) d\psi^2 + \frac{F(y)}{G(y)} dy^2 \right) + F(y)^2 \left( \frac{dx^2}{G(x)} + \frac{G(x)}{F(x)} d\phi^2 \right) \right] \nonumber \eeq
\noindent where
\beq F(\zeta) = 1 -\lambda \zeta,~~~~~~~ G(\zeta) = (1-\zeta^2)(1-\nu \zeta). \nonumber \eeq

\noindent The parameter $R$ indicates the length scale, while $\lambda$ and $\nu$ are dimensionless parameters, with $\nu$ determining the shape of the $S^1 \times S^2$ horizon. 

The coordinates $\phi$ and $\psi$ are periodic with $\Delta \phi = \Delta \Psi = \frac{2\pi\sqrt{1+\lambda}}{1+\nu}$, while the permitted intervals for $x$ and $y$ are determined by the eigenvalues of the metric where the signature is (1,4), the only regions in the $(x,y)$ plane are:
\begin{itemize}
		\item Region $\mathcal{A}_1:~(-1,1) \times (-\infty, -1)$ is asymptotically flat and static. This represents the outer part of the black ring solution, this region can be smoothly connected with $\mathcal{A}_2$ by identifying $y = -\infty$ with $y = \infty$.
		\item Region $\mathcal{A}_2:~(-1,1 ) \times (1/\nu, \infty)$ describes an ergosphere with a limiting surface of stationarity located at $y = \infty$ and a horizon at $y = 1/\nu$.
		\item Region $\mathcal{A}_3:~(-1,1) \times (1/\lambda, 1/\nu)$ is non-stationary and denotes the region below the horizon, the curvature singularity occurs at $y = 1/ \lambda$.
		\item Region $\mathcal{B}:~(-1,1) \times (1,1/\lambda)$ represents the region around a spinning singularity. 
		\item Regions $\mathcal{C}_1:~(1/\lambda, 1/\nu) \times (-1,1)$ and $\mathcal{C}_2:~(1/\nu, \infty) \times (-1, 1)$ are not asymptotically flat, and have no known physical interpretation.
\end{itemize}     

The WANDs for the RBR solution were presented in \cite{brwands} from which it was shown that the Weyl type is generically type ${\bf I}_i$, except on the horizon where it is of type {\bf II}. This was achieved by studying the existence of null vectors $\ell^a = (\alpha,\beta,\gamma, \delta, \epsilon)$ satisfying  
\beq \ell^a \ell^c \ell^{[c} C^{a]~~[d}_{~~bc} \ell^{f]} = 0 \nonumber \eeq
\noindent relative to the coordinate basis. Following the prescription given in equations (22)-(26) in \cite{brwands} a null coframe was built with $\ell$ and $n$ proportional to WANDs, and the Weyl tensor in type ${\bf I}_i$ form was computed. 

By computing the Weyl tensor components for the zeroth iteration of the algorithm, we have explicitly shown  that the null directions stated in \cite{brwands} are indeed WANDs instead of relying on the fact that the type {\bf I} condition \eqref{TypeIcond} is necessary and sufficient for $\ell$ and $n$ to be WANDs. From a computational perspective, this calculation is non-trivial, and it was unachievable using the standard tools provided by Maple or GRTensorII. The coordinate expressions for the Weyl tensor components relative to this coframe are too large to print in the current article, for the interested reader the components and the Maple worksheet used to generate them are included in \cite{Forget}. 

From these components it can be determined that the number of functionally independent invariants at zeroth order is $t_0 = 2$. In the region where the WANDs components are real valued, we may use the Weyl tensor decomposition in table I, and we may infer that the type ${\bf I}_i$ form is affected by a boost and null rotations. At zeroth order, we may entirely fix spatial rotations as the vectors ${\bf \hat{v}}$ and ${\bf \check{v}}$ in table I are orthogonal and lie entirely in the plane spanned by $m^2$ and $m^3$. Therefore, at zeroth order, the dimension of the isotropy group is zero-dimensional, i.e.,  $dim(H_0) =0$.  At the next iteration of the algorithm no new functionally independent invariants appear, so that $t_1 =2$, and the isotropy group is still zero dimensional, $dim(H_1) = 0$. Since $t_0 = t_1 = 2$, and  $dim(H_0) = dim(H_1) = 0$ the algorithm concludes.

\section{Conclusions}

Using the alignment classification, we have introduced the Cartan algorithm as a generalization of the Karlhede algorithm to 5D Lorentzian metrics. Since the alignment classification is applicable to any dimension, this algorithm can be extended to any dimension. While the actual refinement of the Weyl classification in 5D \cite{CHOL2012} exploited the dimensionality of the transverse space to produce simplifications to the Weyl constituents, a similar analysis could be repeated for any fixed dimension.  

As an illustration we have applied the algorithm to three exact black hole solutions. For each spacetime we have generated two discrete sequences summarizing the dimension of the isotropy group and number of functionally independent invariants at each iteration of the algorithm: 
\beq \begin{array}{c|c|c|c} & \text{ Myers-Perry} & \text{Kerr-(A)dS} & \text{RBR} \\ \hline \{ t_q\} & \{2,2,2\} & \{2,2,2\} & \{2,2\} \\ 
\{dim(H_q)\}& \{2,0,0\} & \{2,0,0\} & \{0,0\} \\   \end{array} \nonumber \eeq 
\noindent along with a prescription to generate the required Cartan invariants at each order.  The sequences $\{ t_q\}$ and $\{ dim(H_q)\}$ are discrete invariants, and as such are helpful when showing inequivalence of spacetimes. Unsurprisingly, the sequence $\{ dim(H_q)\}$ shows that the RBR solution is distinct from the Myers-Perry and Kerr-(A)dS black holes. However, the sequences $\{t_q\}$ and $\{dim(H_q)\}$ cannot distinguish the difference between the Myers-Perry and Kerr-(A)dS spacetimes.

In order to do so, one must look at the Cartan invariants generated at zeroth and first order. Choosing $X = C_{0101}$ and $Y=C_{0123}$ as functionally independent zeroth order Cartan invariants, we may rewrite all other invariants in terms of these two invariants, and these expressions will be {\it independent of the choice of coordinates}. For example, $C_{1010;2}$ can be written in terms of $X$ and $Y$ for both metrics: 
\beq \begin{array}{cc} \text {Myers-Perry:} & C_{1010;2} = F_1(X,Y) \\ \text{Kerr-(A)dS:} & C_{1010;2} = F_2(X,Y) \end{array} \nonumber \eeq
with  $F_1 \neq F_2$ when the parameters $a$ and $b$ are non-zero. This proves the metrics are distinct.

The Cartan algorithm may be applied to the identification of a black hole's horizon. We note that all black hole spacetimes are of type {\bf II/D} on the horizon, and that in general a (stationary) black hole spacetime is of type {\bf I} or {\bf G}. We may construct zeroth order SPIs, called discriminants  \cite{DISCRIM, CHDG} which indicate when the Weyl or Ricci tensors are of type {\bf II/D}, and which will vanish on the horizon and are non-vanishing elsewhere. Page and Shoom have shown that it is possible to construct SPIs which detect the event horizon of stationary black holes \cite{PageShoom2015}, this suggests that a first order SPI locates the horizon as well. For the type {\bf II}/{\bf D} spacetimes, like the Myers-Perry or Kerr-AdS spacetimes, we must use a first order invariant to identify the horizon. However, for the rotating black ring spacetime, which is of type ${\bf I}_i$ generally, both the type {\bf II}/{\bf D} zeroth order invariants and the first order invariants vanish on the horizon. We can use the Cartan algorithm directly to produce Cartan invariants that detect the horizon; this has been illustrated for several 4D and 5D black hole solutions where SPIs and Cartan invariants were compared for horizon detection \cite{GANG}. 

Motivated by the well understood calculation of gravitational wave signals in the theoretical modelling of 4D sources in GR,  gravitational wave extraction from numerical simulations of rapidly spinning objects in higher dimensions has been studied \cite{Cook}. This method utilizes a particular frame and employs projections of the Weyl tensor components to calculate the gravitational waves in spacetimes with rotational symmetry. From the perspective of the Cartan algorithm this suggests the use of Cartan invariants. In terms of computability, the Cartan invariants have an advantage over SPIs since the Cartan invariants are easier to compute than the related SPI. For example, by  relaxing the condition for using a coframe based on WANDs allows for the Cartan algorithm to produce Cartan invariants that detect the horizon \cite{AM2016}. We shall pursue this in future work \cite{ADA1}.

\newpage

\section*{Acknowledgements}  
 
This work was supported through the Research Council of Norway, Toppforsk grant no. 250367: Pseudo-
Riemannian Geometry and Polynomial Curvature Invariants: Classification, Characterisation and Applications (D.M.), and NSERC (A.A.C., A.F.). 

\providecommand{\href}[2]{#2}
\begingroup\raggedright
{}
\end{document}